# Crowdfunding for Equitable EV Charging Infrastructure

Abdolmajid Erfani[1], Qingbin Cui[*2] and Patrick DeCorla-Souza[3]

**ABSTRACT**

The transportation sector significantly contributes to greenhouse gas emissions, highlighting the need to transition to Electric Vehicles (EVs) to reduce fossil fuel dependence and combat climate change. The US government has set ambitious targets for 2030, aiming for half of all new vehicles sold to be zero-emissions. Expanding EV charging stations is crucial for this transition, but social equity presents a significant challenge. The Justice40 program mandates that at least 40% of benefits be allocated to disadvantaged communities, ensuring they benefit from federal investments. Given the current concentration of EV ownership in affluent areas, merely installing charging stations in disadvantaged neighborhoods may not suffice. This article explores crowdfunding as a novel method to finance EV charging infrastructure, engaging, and empowering underserved communities. The paper concludes with a hypothetical case showing financing benefits for disadvantaged communities, exploring crowdfunding variations, and scaling to develop equitable EV charging networks.

## 1. Introduction

[1] Assistant Professor, Michigan Technological University, Houghton, MI, USA, 49931, aerfani@mtu.edu

[2] Professor, University of Maryland, College Park, MD, USA, 20742, cui@umd.edu, (Corresponding Author)

[3] P3 Program Manager, Federal Highway Administration, Washington, D.C., USA, 20590, Patrick.DeCorla-Souza@dot.gov

Transportation is the largest contributor to greenhouse gas (GHG) emissions in the U.S., contributing about 29 percent of total GHG emissions (Albuquerque et al., 2020; Solaymani, 2019; EPA, 2023). Therefore, transitioning to Electric Vehicles (EVs) can help mitigate air pollution and climate change, as well as decrease dependence on fossil fuels and promote energy independence (Sun et al., 2023; Zhang & Fujimori, 2020). To make EVs accessible to all Americans for local and long-distance trips, the U.S. government plans to make half of all new vehicles sold in 2030 zero-emissions vehicles, as well as build a network of 500,000 charging stations that are convenient and equitable (USDOT, 2023). With the new Infrastructure Investment and Jobs Act, $7.5 billion has been allocated to EV charging stations and alternative fueling to demonstrate government support for the conversion to EVs. However, significant obstacles still stand in the way of the widespread adoption of EVs, from their high cost to inequitable EV charging stations placement (Lebrouhi et al. 2021). Although there has been some significant support and growth in the EV charging infrastructure development, the market for EV charging remains uncertain. For example, the top 100 U.S. metropolitan areas are expected to require more than 195,000 non-residential EV charging points by 2025, almost four times what they had at the end of 2017 (Nicholas et al., 2019). A growing trend toward EV adoption will also increase this need.

Due to the current paradigm shift towards automation, electrification, and shared infrastructure in transportation (Sperling 2018), appropriate allocation of EV charging infrastructure is imperative. Providing electric mobility solutions to diverse communities can lead to a more inclusive environmental movement, with the advantages of reduced emissions and sustainable living shared by everyone (Pan et al. 2021). This underscores the significance of equitable planning for EV charging infrastructure (Li et al. 2022), especially as the United States embarks on a new era of EV planning bolstered by substantial governmental grant support (Carlton and Sultana 2024).

Current EV planning prioritizes optimizing locations based on utility complexities and market demand, yet there's an increasing demand for incorporating a socially aware perspective (Mandolakani and Singleton 2024; Moriarty 2022). A study by Roy and Law (2022) highlights the spatial disparities in access to EV charging stations, advocating for the integration of social equity with other factors in the development of these stations.

There is such a need for equitable infrastructure development that the US White House issued Executive Order 13985 and Executive Order 2023 "Further Advancing Racial Equity and Support for Underserved Communities" (The White House, 2022). The Justice40 Initiative allocates 40% of federal funding to sustainable transportation for underserved communities (Yang et al., 2021). There are several challenges associated with building EV charging infrastructure that meets demand and ensures Justice40 compliance. While the Justice 40 initiative aims to direct 40% of federal investments towards disadvantaged communities, since EV ownership is currently concentrated in relatively wealthy areas, installing charging stations alone may not be sufficient to meet the overall demand and fulfill Justice40's goals (Caulfield et al., 2022; Bonsu, 2021; Adepetu & Keshav, 2017). Placing charging infrastructure solely in affluent neighborhoods may perpetuate existing inequities and limit access for disadvantaged communities. On the other hand, disadvantaged communities often face financial constraints, making it difficult for them to purchase and own EVs. There is therefore a need for a solution that addresses both the current need for EV charging stations and the equitable allocation of federal funding. The purpose of this paper is to study an EV crowdfunding program as a public engagement approach aimed at addressing the challenges through meeting EV charging demand and ensuring compliance with Justice40.

In addition, while recent attention has been drawn to equity analysis in transportation projects, studies on EV charging networks development are limited. Several previous studies have examined fairness in accessibility. Access to healthcare (Lane et al., 2017), EV charging stations (Nazari-Heris et al., 2022), parks (Chen et al., 2023), and other infrastructure (Erfani et al. 2024) is facilitated by a variety of spatial and temporal accessibility measures (Geurs et al., 2016; Wang et al. 2024). It is important to remember that the equity issue in the EV market consists of two components, namely (a) the ownership of EVs and (b) the provision of EV charging infrastructure. Since EV adoption is increasing, balancing these two parts is essential. In this regard, the paper discusses how to implement EV crowdfunding program and demonstrate a hypothetical case. Finally, the paper discusses future research directions and the variations of the crowdfunding models.

## 2. Literature review

The literature review of this study interconnects three multidisciplinary areas of consideration in the process of planning for developing EV charging networks. These areas encompass current technical and demand considerations, relevant equity considerations in access to infrastructures such as EV charging stations. Additionally, the review examines how crowdfunding has been explored in other domains. The integration of these three areas is closely linked to the proposed approach in this study for better engagement of underserved communities in EV charging investment.

### 2.1 EV charging station location planning

Due to the availability of newer EV models with greater driving ranges, a variety of models to choose from, more affordable options, and better public incentives, the U.S. is well positioned for increasing its EV adoption rate (Ledna et al., 2022). Increasing EV adoption calls for public access to EV charging infrastructure that supports both slow and fast charging, which requires major planning to determine locations for the network. As a result, the problem of charging station location has become a research hotspot lately. In most cases, the location determination problem is framed in terms of technical, economic, and user acceptance issues (Metais et al., 2022). Technically, since public charging infrastructure uses energy from the grid, it is important to consider the constraints associated with power grid operation and energy storage (Amry et al., 2023). Aside from that, since EVs have a different refueling behavior and their batteries can charge while they aren't in use for mobility purposes, it is important to select locations that are suitable for drivers, such as shopping malls, recreation facilities, public parking facilities, and other public spaces. As a result, user acceptance and their needs play a major role in the design of EV charging networks location (Metais et al., 2022; Yi et al., 2019). Because charging rates and costs vary, more expensive infrastructure should result in higher charging rates for users. In addition to penalizing consumers, a poor choice of EV supply equipment also penalizes operators whose return on investment is better if a charging station is tailored to local needs. Therefore, cost-time-effective location determination solutions become crucial (Ahmad et al., 2022; Li et al., 2016; Zhao et al. 2020).

In terms of location modeling, a relatively new branch of the literature addresses location determination with flow-based demand, as opposed to classical facility location models. In this approach, due to the limited driving range of EVs, drivers who travel longer distances, usually exceeding their driving range, will have to recharge their batteries. Therefore, recharging demand can be modelled based on origin-destination (OD) of trips (Kchaou-Boujelben, 2021). To solve and optimize the charging problem, a variety of heuristics, approximate, and exact methods have been used. While the literature on the technical side of location optimization is extensive, there are other decision variables that contribute to this problem such as technology choice, user demand, local regulations, incentives, operational considerations, and more importantly social equity (Panah et al., 2022; Metais et al., 2022; Li et al., 2016; Kchaou-Boujelben, 2021).

**2.2 Equitable infrastructure development**

Engineers' diversity, equity, and inclusion (DEI) studies primarily examine the internal engineering workforce of an organization (Hickey et al., 2022; Leuenberger et al., 2022; Erfani et al., 2023), infrastructure siting, design decisions, the impact of individual construction projects (Gandy et al., 2023), or the social costs associated with failures and disruptions (Coleman et al., 2020; Li et al. 2023). A growing emphasis has been placed on integrating social equity into transportation planning and investment (Soliz et al., 2023). By ensuring equal access to resources and opportunities for all, regardless of backgrounds, characteristics, or circumstances, equity can be attained. It is crucial to ensure infrastructure equity to eliminate social and economic disparities, promote inclusiveness, and contribute to sustainable development. Vertical equity and horizontal equity are the most discussed types of equity in transportation literature. According to a horizontal perspective, equity refers to the distribution of resources and opportunities equally (or "fairly") to individuals and classes considered equally capable and needy (Delbosc & Currie, 2011). Consequently, horizontal equity involves everyone bearing the same costs, receiving the same benefits, and having the same opportunities. Vertical equity, on the other hand, refers to the distribution of resources among people and groups with differing abilities and needs. Vertical equity eliminates inequalities between socioeconomic groups and individuals by promoting unequal distributions of resources (Nahmias-Biran et al., 2014).

In the context of EV charging infrastructure, an equitable distribution of public, mobile, and residential types of EV charging infrastructure have been studied (Caulfield et al., 2022; Nazari-Heris et al., 2022; Carlton & Sultana, 2022; El Hachem and De Giovanni, 2019). In addition to focusing on inequity in the distribution of EV charging infrastructure, it is imperative to take into consideration the effect of access to EV charging (Canepa et al., 2019). In studies looking at the relationship between charging infrastructure distribution and socio-demographic data, it was found that population density was not associated with EV charging stations (Khan et al., 2022; Law & Roy, 2021), but was correlated with median household income (Carlton & Sultana, 2022; Khan et al., 2022; Law & Roy, 2021; Min & Lee, 2020), age (Law & Roy, 2021), white-identified population (Khan et al., 2022), local EV adoption (Borlaug et al. 2023), education level (Pan et al. 2024), and highway presence (Khan et al., 2022; Hsu & Fingerman, 2021). While mobile charging stations have gained attention because of their flexibility for public networks, charging networks should be convenient, equitable, and affordable for users, but should also provide providers with a return on investment (Hopkins et al., 2023). As a result, it is imperative to develop innovative approaches to balance the needs of charging infrastructure for EVs with the needs of social equity, such as Justice40 in the U.S.

The U.S. government's Justice40 Initiative mandates that at least 40% of benefits from federal investments in climate change, clean energy, clean transit, and related sectors be directed to disadvantaged communities burdened by pollution (The White House, 2022). Justice40 underscores the inseparable links between climate action, environmental justice, and social equity, echoing the perspectives of leading scholars in the field (Faber, 2008; Gilio-Whitaker, 2019; Kojola & Pellow, 2021). Creating maps and tools (such as the Climate and Economic Justice Screening Tool, USDOT Equitable Transportation Community Explorer), along with action plans and definitions for disadvantaged communities, plays a crucial role in channeling resources and benefits to these areas. Although the process of identifying and tackling environmental pollution through Justice40 continues to be a subject of active debate (Wang et al. 2023), it represents a significant move towards a more equitable distribution of resources across communities (Sotolongo 2023).

**2.3 Crowdfunding**

Crowdfunding refers to the practice of raising funds from a large number of individuals or organizations through online platforms to support different types of projects. It operates on the principle of collective financing, where individuals contribute small amounts of money to collectively meet the funding needs of a project (Farajian et al. 2015; Butticè & Ughetto, 2021). The key benefit of crowdfunding in infrastructure is that it allows project sponsors to access a diverse pool of potential funders, including community members, local businesses, and other interested parties. This opens up new avenues for financing that may not be available through traditional sources such as government grants or bank loans. By tapping into the power of the crowd, project sponsors can generate a sense of ownership and engagement within the community, fostering a greater sense of support and involvement in the project's success (Gerber et al., 2012).

Depending on the funding purpose and investment method, crowdfunding can be classified as donation-based, reward-based, equity-based, or lending-based (Best et al., 2013). Donation-based crowdfunding has been used by non-governmental organizations (NGOs) for more than a decade to raise funds for their missions and projects (DeBuysere et al., 2012). Entrepreneurs who are trying to raise money for a campaign often use reward-based crowdfunding. Customers-oriented products and services are increasingly sold through reward-based crowdfunding. By matching lenders with borrowers through online platforms, lending-based crowdfunding provides unsecured loans. A crowdfunding platform sets the interest rate, which is typically higher than saving rates and lower than traditional loans. Through an online platform, equity-based crowdfunding allows private businesses to offer securities to the public for sale (Lam et al., 2016;

Vulkan et al., 2016). In some classifications, reward-based and donation-based crowdfunding is referred to as "community crowdfunding", while equity-based and lender-based crowdfunding is referred to as "financial return crowdfunding" or "investment crowdfunding" (Kirby & Worner; 2014).

In addition to its business applications, crowdfunding has also been applied to civil and infrastructure projects. For example, in light of crowdfunding's success on platforms like Kickstarter, high-tech entrepreneurs are investigating how crowdfunding can be used in civic projects, especially given government budgetary constraints. One example of a crowdfunded civic project that succeeded is The Low Line in New York City, which transformed an abandoned train station into an underground park. In 2012, it raised $150,000 on Kickstarter (Gray, et al., 2013). Crowdfunding has also been shown to enhance traditional public-private partnership (PPP) delivery models (Farajian et al., 2015; Farajian & Ross, 2016). Several research studies have shown that crowdfunding can contribute to the delivery of renewable energy projects and a variety of other types of projects as well (Lam & Law, 2016; Lu et al., 2018). It has been studied from a profit perspective that using crowdfunding in infrastructure development can have an effect equivalent to a 20% subsidy (Zhu et al., 2017). However, another advantage of crowdfunding to promote underserved communities has not been studied. As a result, it is worth considering crowdfunding for EV charging infrastructure development not just for financing but also for empowering and engaging underserved communities with federal investments. The purpose of this study is to demonstrate how crowdfunding models and their variations can be used to address current EV charging station needs while complying with Justice40 and equity considerations.

## 3. RESEARCH METHODOLOGY

### 3.1 EV Crowdfunding program design

EV infrastructure crowdfunding represents an innovative and unique approach to promoting equity and inclusion in the construction of charging infrastructure (see Figure 1). Figure 1 illustrates the distinction between the crowdfunding method (Crowd-funded PPP) and the conventional PPP approach. Reviewing state and local agency examples of PPP utilization reveals that these projects rely on charging network revenue. Private parties typically assume upfront risks and responsibilities for design, construction, financing, operation, and maintenance, with revenue linked to location demand. Traditional PPP models make it difficult to benefit disadvantaged communities while ensuring high charging station revenue.

<Insert Figure 1 here>

**Traditional PPP versus Crowd-funded PPP (CPPP)**

PPP financing involves a collaborative agreement between government entities and private sector companies to fund, build, and operate projects, such as public transportation networks, parks, and infrastructure, sharing both the investments and the risks (Papajohn et al. 2011). Although the PPP model is well-established, it offers limited to no opportunities for institutional or individual investors to hold a direct equity stake in these projects (Farajian and Ross 2016). Instead, within this model, the opportunity for direct equity is mainly confined to large infrastructure funds and developers. Crowd-funded PPP maintains a similar structure to traditional PPP but introduces added flexibility by allowing both institutional and individual investors (specifically focusing on disadvantaged communities in the context of EV usage) to participate as direct equity contributors (Figure 1). This crowd-funded approach evolves the existing PPP model, generating value that extends beyond mere financial advantages (Farajian et al. 2015).

Unlike typical crowdfunding in infrastructure projects, this specific crowdfunding program focuses on engaging and empowering the communities that have been historically underserved. One distinguishing feature of this crowdfunding initiative is the offering of charging station ownership shares to community members at high discounts. By providing ownership opportunities, community members become active stakeholders in the project, aligning their interests with the success and sustainability of the EV infrastructure. This approach not only fosters a sense of ownership and pride but also ensures that the benefits of the charging stations directly flow back to the community. A key advantage is aligning with the Justice40 initiative's requirement, ensuring that 40% of federal funds and grants are allocated directly to support disadvantaged communities. This is upheld even if those communities do not own EVs or have EV charging stations installed in their neighborhoods. Unlike traditional PPP models, where direct-equity opportunities are mainly restricted to large infrastructure developers, CPPP broadens the scope by enabling the participation of individuals, specifically from selected and verified disadvantaged communities, in the proposed EV program.

Furthermore, the discounts offered on ownership shares make participation more accessible and attractive to individuals in disadvantaged communities who may have limited financial resources. This aspect of the crowdfunding initiative helps bridge the affordability gap and encourages community members to invest in clean transportation infrastructure that directly benefits their neighborhoods. By targeting disadvantaged communities, this crowdfunding program addresses the disparities in EV ownership and charging infrastructure access. It creates an avenue for community members to actively participate in the transition to electric mobility and enjoy the benefits of reduced emissions, cleaner air quality, and increased transportation equity.

The following structure is considered in order to implement the proposed EV crowdfunding program. This is a practice-oriented recommendation that includes details on the set-up and reporting of the platform. In addition, we will demonstrate the calculation of this equity based crowdfunding model on a hypothetical EV charging development project. The following steps are recommended for a successful crowdfunding program (Davis and Cartwright 2019; Forbes and Schaefer 2017).

- **Needs Assessment and Feasibility Study:** Conduct a thorough needs assessment to identify target disadvantaged communities and assess their readiness and interest in participating in the crowdfunding program. Evaluate the feasibility of implementing the program, considering factors such as local regulations, available charging infrastructure, and community support (Logue and Grimes 2022). This step is crucial for studying the economic behavior of underserved communities (Dixit and Nalebuff, 2008) to ensure their interest and participation in the program, thereby contributing to its success.

- **Program Development:** Develop a comprehensive program framework that outlines the goals, objectives, and guidelines for the crowdfunding pilot. Define the ownership structure, including the discount rates and ownership share options to be offered to community members. Establish clear eligibility criteria and guidelines for community participation in the crowdfunding program. Design an engagement strategy to raise awareness and foster community buy-in, including outreach efforts, community meetings, and educational campaigns (Samarah, and Alkhatib 2020).

- **Platform Selection and Setup:** Identify and select a suitable crowdfunding platform that aligns with the goals and requirements of the program. Set up the crowdfunding platform, ensuring it allows for the creation of ownership shares, secure transactions, and transparent reporting. Customize the platform to incorporate program-specific features and branding (Gooch et al. 2020).

- **Community Engagement and Promotion:** Implement an outreach campaign to inform and engage community members about the crowdfunding program. Organize community meetings, workshops, and webinars to educate participants about the benefits of EV infrastructure and ownership opportunities. Collaborate with local organizations, community leaders, and influencers to promote the crowdfunding initiative (Efrat et al. 2020).

- **Evaluation and Selection Process:** Develop a transparent and fair evaluation process to review and select community members for ownership shares. Establish criteria for selection, such as residency, financial need, and community involvement. Ensure the selection process is well-documented, impartial, and communicated clearly to participants (Forbes and Schaefer 2017).

- **Implementation and Monitoring:** Facilitate the installation and operation of charging stations, utilizing the funds raised through the crowdfunding pilot. Regularly monitor and evaluate the progress and impact of the EV infrastructure in disadvantaged communities. Collect feedback from community members and stakeholders to identify areas for improvement and address any challenges (Mastrangelo et al. 2020).

- **Reporting and Documentation:** Maintain comprehensive records of the crowdfunding program, including financial data, ownership agreements, and community participation (Nan et al. 2019). Generate reports on the program's outcomes, highlighting the number of charging stations installed, community involvement, and socioeconomic benefits.

- **Scaling and Expansion:** Assess the success and lessons learned from the program to determine the potential for scaling up and expanding the initiative to additional communities.

Seek partnerships and explore additional funding sources to support the growth and sustainability of the crowdfunding program (Lam and Law 2016).

## 3.2 Hypothetical example

**Deterministic Analysis: Crowdfunding Conceptual Framework**

The purpose of this section is to demonstrate in a hypothetical example the financing process and benefit calculations for disadvantaged communities that participate in the crowdfunding program for building EV charging networks. In this case, Montgomery County in Maryland will be used as an example. To encourage residents and businesses to switch to eco-friendly transportation options, Montgomery County has implemented numerous incentives, infrastructure improvements, and awareness campaigns. As part of its Climate Action Plan (Montgomery County, 2021), Montgomery County has set an ambitious target of reducing carbon emissions by 80 percent by 2027 and 100 percent by 2035. As part of those reductions, the County officials are focusing on the transportation sector, which contributes 42 percent of the county's climate-changing emissions. The numbers and examples used in this section are intended to demonstrate the challenges this county faces in applying for federal funding programs like Charging and Fueling Infrastructure (CFI) discretionary grant program and working with the current trend of electric car usage and ambitious climate goals.

As shown in Figure 2, EV charging network placement has been selected based on neighborhood EV registrations and projected utilization rates across the county. Additionally, the figure shows historically disadvantaged communities (HDC) according to US DOT's Equitable Transportation Community (ETC) Explorer. The ETC tool recognizes disadvantaged communities in the following five component areas: transportation insecurity, climate and disaster risk burden, environmental burden, health vulnerability, and social vulnerability. The tool examines the cumulative burden disadvantaged communities face as a result of underinvestment in transportation. Using newly available 2020 Census Tracts data, the tool adds additional indicators reflecting a lack of investment in transportation (USDOT, 2023). All EV network areas do not lie in HDC areas which challenge Justise40 compliance. This is a common issue in establishing and building EV charging infrastructure, where demand and EV adoption do not necessarily align with underserved communities' neighborhoods.

<Insert Figure 2 here>

For the purpose of illustration, we assume that 40 Level 3 Superchargers (350kW) and 60 Level 2 chargers are installed in the highlighted areas of Figure 2. While other research estimates cost items and revenue of EV charging networks based on usage rate, energy price, and other uncertainties related to this topic (Nicholas, 2019; Gamage and Jenn 2023), we have included cost items and revenue projections based on average data available for EV networks for demonstration purposes (Nicholas 2019; Satterfield and Nigro 2020). Table 1 provides details on the funds required for this project, which total $10 M.

<Insert Table 1 here>

Although the estimates are for demonstration purposes in this hypothetical example, reasonable estimates have been included (Nigro 2019). Cost items include the installation of type 3 and type 2 chargers as well as other indirect costs related to the entire project, including design, traffic management, public engagement, and operation. In alignment with the traditional PPP model, operational costs are primarily the responsibility of private investors. However, with the addition of government subsidies, a portion of these operational expenses can be offset. For instance, in our example, we employ a 10-year lifecycle analysis for the chargers, during which the CFI grant program contributes to five years of operational costs. In order to implement the EV crowdfunding program and calculate further benefits, the following financing structure is considered. The CFI program requires a minimum 20% matching fund for the project, so the county is considering receiving $8 M from the federal government out of 10 million dollars. The remaining $2 M is composed of $1.5 M from private investors and 500,000$ for HDC ownership on crowdfunding program. For selected HDC communities to participate in the crowdfunding program, 5000 shares with high discount rates are proposed at a $100 price. In the proposed crowdfunding program, a process will be created to review and select community members who apply to own shares, with a maximum limit on the number of shares that can be purchased by a single individual. An equity structure is proposed for the crowdfunding program that would give 50% ownership to the operator and 50% to HDC communities (Figure 1). Accordingly, 50% of net project revenue (after any expenditures on operations and maintenance) will go to the operator and 50% to the owners through the crowdfunding program. These numbers are for illustration purposes only; different sets of numbers could be applied to evaluate the impact on the outcome.

Table 2 demonstrate the return on investment according to project revenue and equity structure to compare a traditional PPP model with the CPPP model. In the first five years of the lifecycle of the project, assuming that operating costs are covered by the CFI program, the HDC community will receive $72 per share per year. For the remainder of the project lifecycle, each crowdfunding program share will produce $22 per share annually.

< Insert Table 2 here>

From the perspective of the HDC community, investment cashflow includes a $100 payment in Year 0, $72 received each year over the first five years of operations, and $22 each year in the subsequent five years. An analysis of this investment shows an internal rate of return

(IRR) of 68%, making it a potential attractive investment for HDC. Although the financial analysis in Table 2 accounts solely for direct revenue from electricity charges, a benefit-cost analysis could encompass additional advantages, including carbon dioxide emission reductions, climate change mitigation benefits, and enhancements in energy security. Furthermore, with the rising popularity of EVs and the anticipated increase in demand for EV charging, revenue is expected to grow throughout the project lifecycle. Nonetheless, a consistent revenue stream has been assumed for demonstration purposes. The goal is not to prove financial benefits, but to demonstrate how crowdfunding can serve as a financing tool to empower underserved communities and ensure compliance with Justice40 for equitable infrastructure development.

The analysis adopted a deterministic method to illustrate the crowdfunding concept to finance EV charging stations in alignment with Justice40 requirements. The anticipated revenue and ultimate IRR for HDC communities are influenced by various elements, including charging locations, type of chargers, as well as factors like share price, discount rates, size of the crowdfunding program, and the percentage of matching funds. To accurately determine a guaranteed IRR that would appeal to HDC investors for implementation, a more thorough and extensive analysis is necessary to address all these complexities.

## 4. Discussion: Crowdfunding program variations

Many US federal funding programs allocate a significant portion of their resources to projects that benefit disadvantaged and underserved communities. Setting criteria that prioritize projects in low-income and historically marginalized areas, such as the ETC Explorer and Climate and Economic Justice Screening (CEJST) Tool, can help direct funding where it is most needed. When it comes to projects like EV charging infrastructure, where demand may be concentrated in wealthier areas, it can indeed be challenging to ensure compliance by providing 40% of benefits to disadvantaged communities. Crowdfunding can be an effective strategy to promote equity and ensure Justice40 goals are met by providing benefits to disadvantaged communities who own the EV charging network, rather than the vehicles. The model emphasizes empowering communities through direct involvement in the charging infrastructure, which fosters a sense of ownership and control over one of the most important elements of sustainable transportation. Even though this study focused on introducing equity-based crowdfunding to make ownership shares more financially accessible for community members with limited resources to get involved, there may be other variations and strategies to address this issue as well.

Rather than offering shares to individuals, a crowdfunding program could offer community shares, which provide participants in the community investment with non-monetary benefits, instead of financial returns. Several benefits could be offered, such as priority access to the charging infrastructure, discounts on charging rates, or other incentives that encourage community involvement (Hopkins et al. 2023). Individuals who participate in crowdfunding programs may be offered additional incentives, such as discounts on EV purchases or lease options, to encourage them to adopt EVs in underserved communities.

To consider the risks in the EV market and entice more investors from disadvantaged communities, a synergistic approach can be developed that connects EV ownership with the ownership of charging infrastructure. This could be expanded to a rebate program for HDC communities. To promote the adoption of EVs while addressing social and economic disparities, rebates can be offered to disadvantaged communities to help them purchase EVs. There is room to add a donation matching program for HDC communities, which will increase partnership and benefits. Corporations and foundations can partner with community organizations to establish donor-matching programs. By matching every dollar raised through crowdfunding, a partner organization effectively doubles the amount available for the development of EV charging infrastructure.

Finally, EV charging infrastructure projects can be financed with green bonds to attract socially responsible investors. Funds raised from the bond issuance can be used to build and operate charging stations in underserved areas. The advantages of each of these variations can be tailored to meet the specific needs and goals of each community and region. Exploring and implementing innovative crowdfunding models for EV charging infrastructure can enhance community engagement, promote social equity, and hasten the shift towards a sustainable mobility future.

This study delved into the use of the crowdfunding model, offering investment discounts to selected and verified disadvantaged communities to achieve Justice40 compliance. However, crowdfunding approach opens the possibility for all interested investors to contribute, enhancing flexibility and extending beyond traditional large infrastructure investment sources (Zhu et al. 2017). Nevertheless, while this strategy presents clear benefits, it also introduces some challenges. Prior attempts to incorporate crowdfunding into civil infrastructure financing have identified obstacles such as increased complexity, administrative and accounting difficulties, and issues with managing confidential information from external sources (Farajian et al. 2015). Implementing a crowdfunded model would add further complexities to contract documentation, procurement strategies, and project management, necessitating careful oversight and additional resources (Pranata et al. 2022).

Moreover, given the current uncertain landscape of the EV market, risk allocation plays a pivotal role in this crowdfunding program. The allocation of risk should be influenced by the structure of the payment schedule and order in the contract. Contractual terms need to prioritize an equitable approach, especially detailing the payment process for scenarios of diminished revenue. Although this aspect is crucial, it necessitates a thorough, case-by-case assessment that falls beyond the scope of this study.

## 5. Conclusion

EVs are crucial to combating climate change, reducing greenhouse gas emissions, improving air quality, and achieving a sustainable and environmentally responsible transportation system. In prior studies, it was found that the current distribution of EV charging stations is not equitably distributed across the US, since EV charging stations were not correlated with population density, but rather with median household income, age, and racial background. Further, the current concentration of EV owners in wealthy areas makes it difficult for federal money to support grant programs that develop EV charging networks due to Justice40 constraints. In accordance with Justice40, at least 40% of investment benefits must go to HDC communities. Yet, members of HDC communities that do not own an EV might not benefit from the network if it is only located in their neighborhood. Therefore, to address current demands for EV charging networks and to comply with Justise40, this article proposes an EV crowdfunding program. Through this crowdfunding program, disadvantaged communities can invest in the development of EV charging stations. By offering discounted ownership shares to community members with limited resources, it becomes more financially accessible. In this way, they become active stakeholders in the project and benefit from the charging network financially or otherwise.

A hypothetical example of a crowdfunding program in Montgomery County, Maryland was used to demonstrate the program structure and benefit calculations. Despite the limitations of our assumptions and revenue calculations in a deterministic approach considering all uncertainties in location figuration, crowdfunding program size, price per share, the potential of this crowdfunding program to engage and benefit the HDC community is demonstrated. The purpose of this study was to demonstrate how crowdfunding can go beyond being a mere financing avenue for infrastructure projects to actively engage underserved communities. It is necessary to conduct further research and other efforts to develop detailed program setups, community detection techniques, outreach programs, and to implement this crowdfunding program in practice. The potential of innovative strategies, including rebates and donor-matching, to enhance this crowdfunding program has been discussed. Additional research is required to assess the effectiveness and appeal of these methods.

Given the volatile investment climate for EV charging infrastructure, the numbers and analysis in our study are based on certain assumptions. The primary aim is to explore how a crowdfunding model can engage disadvantaged communities in ownership of charging stations, aligning with Justice40 requirements. Launching the program hinges on several crucial elements, including the participation of disadvantaged communities willing to invest despite market uncertainties. Federal grants and subsidies that alleviate installation and operational expenses play a pivotal role in driving the EV market forward amidst these challenges. Future research should delve into the implementation, effectiveness, and critical determinants of the proposed model's performance, examining aspects such as location configuration, type of charging stations, size of the crowdfunding program, price per share, and other significant factors.

**ACKNOWLEDGEMENT**

Patrick DeCorla-Souza co-authored this paper in his personal capacity. The views expressed in this paper are those of authors and not necessarily the views of the U.S. Department of Transportation (USDOT) or the Federal Highway Administration (FHWA).

**CRediT authorship contribution statement**

**Abdolmajid Erfani:** Data collection, analysis and interpretation of results, Writing – review & editing. **Qingbin Cui:** Conceptualization, Writing – review & editing. **Patrick DeCorla-Souza:** Conceptualization, Writing – review & editing

**REFERENCES**

Adepetu, A., & Keshav, S. (2017). The relative importance of price and driving range on electric vehicle adoption: Los Angeles case study. *Transportation*, 44, 353-373.

Ahmad, F., Iqbal, A., Ashraf, I., & Marzband, M. (2022). Optimal location of electric vehicle charging station and its impact on distribution network: A review. *Energy Reports*, 8, 2314-2333.

Albuquerque, F. D., Maraqa, M. A., Chowdhury, R., Mauga, T., & Alzard, M. (2020). Greenhouse gas emissions associated with road transport projects: current status, benchmarking, and assessment tools. *Transportation Research Procedia*, 48, 2018-2030.

Amry, Y., Elbouchikhi, E., Le Gall, F., Ghogho, M., & El Hani, S. (2023). Optimal sizing and energy management strategy for EV workplace charging station considering PV and flywheel energy storage system. *Journal of Energy Storage*, 62, 106937.

Best, J., Lambkin, A., Neiss, S., Raymond, S., & Swart, R. (2013). Crowdfunding's potential for the developing world. InfoDev. Washington DC, 1.

Bonsu, N. O. (2021). Net-zero emission vehicles shift and equitable ownership in low-income households and communities: why responsible and circularity business models are essential. Discover Sustainability, 2, 1-9.

Borlaug, B., Yang, F., Pritchard, E., Wood, E., & Gonder, J. (2023). Public electric vehicle charging station utilization in the United States. *Transportation Research Part D: Transport and Environment*, 114, 103564.

Butticè, V., & Ughetto, E. (2021). What, where, who, and how? A bibliometric study of crowdfunding research. *IEEE Transactions on Engineering Management*.

Canepa, K., Hardman, S., & Tal, G. (2019). An early look at plug-in electric vehicle adoption in disadvantaged communities in California. *Transport Policy*, 78, 19-30.

Carlton, G. J., & Sultana, S. (2022). Electric vehicle charging station accessibility and land use clustering: A case study of the Chicago region. *Journal of Urban Mobility*, 2, 100019.

Carlton, G. J., & Sultana, S. (2024). Electric vehicle charging equity and accessibility: A comprehensive United States policy analysis. *Transportation Research Part D: Transport and Environment*, *129*, 104123.

Caulfield, B., Furszyfer, D., Stefaniec, A., & Foley, A. (2022). Measuring the equity impacts of government subsidies for electric vehicles. *Energy*, 248, 123588.

Chen, J., Kinoshita, T., Li, H., Luo, S., Su, D., Yang, X., & Hu, Y. (2023). Toward green equity: An extensive study on urban form and green space equity for shrinking cities. *Sustainable Cities and Society*, *90*, 104395.

Coleman, N., Esmalian, A., & Mostafavi, A. (2020). Equitable resilience in infrastructure systems: Empirical assessment of disparities in hardship experiences of vulnerable populations during service disruptions. *Natural Hazards Review*, 21(4), 04020034.

Davis, M., & Cartwright, L. (2019). Financing for society: Assessing the suitability of crowdfunding for the public sector. Report. University of Leeds. https://doi.org/10.5518/100/7

DeBuysere K, Gajda O, Kleverlaan R, Marom D, Klaes M. (2012). A framework for European crowdfunding. European Crowdfunding Network (ECN).

Delbosc, A., & Currie, G. (2011). Using Lorenz curves to assess public transport equity. *Journal of Transport Geography*, 19(6), 1252-1259.

Dixit, A. K., & Nalebuff, B. (2008). The art of strategy: a game theorist's guide to success in business & life. WW Norton & Company.

Erfani, A., Hickey, P. J., & Cui, Q. (2023). Likeability versus Competence Dilemma: Text Mining Approach Using LinkedIn Data. *Journal of Management in Engineering*, 39(3), 04023013.

Erfani, A., Mahmoudi, J., & Cui, Q. (2024). Measuring Social Equity in Pavement Conditions Using Big Data. In *Construction Research Congress 2024* (pp. 23-32).

El Hachem, W., & De Giovanni, P. (2019). Accelerating the transition to alternative fuel vehicles through a Distributive Justice perspective. *Transportation Research Part D: Transport and Environment*, 75, 72-86.

Efrat, K., Gilboa, S., & Sherman, A. (2020). The role of supporter engagement in enhancing crowdfunding success. *Baltic Journal of Management*, 15(2), 199-213.

Faber, D. (2008). *Capitalizing on environmental injustice: The polluter-industrial complex in the age of globalization*. Rowman & Littlefield Publishers.

Farajian, M., & Ross, B. (2016). Crowd Financing for Public–Private Partnerships in the United States: How Would It Work? *Transportation Research Record*, 2597(1), 44-51.

Farajian, M., Lauzon, A. J., & Cui, Q. (2015). Introduction to a crowdfunded public–private partnership model in the United States: policy review on crowdfund investing. *Transportation Research Record*, 2530(1), 36-43.

Forbes, H., & Schaefer, D. (2017). Guidelines for successful crowdfunding. *Procedia cirp*, *60*, 398-403.

Gamage, T., Tal, G., & Jenn, A. T. (2023). The costs and challenges of installing corridor DC Fast Chargers in California. *Case Studies on Transport Policy*, 11, 100969.

Gandy, C. A., Armanios, D. E., & Samaras, C. (2023). Social Equity of Bridge Management. *Journal of Management in Engineering*, 39(5), 04023027.

Gerber, E. M., Hui, J. S., & Kuo, P. Y. (2012). Crowdfunding: Why people are motivated to post and fund projects on crowdfunding platforms. In Proceedings of the international workshop on design, influence, and social technologies: techniques, impacts and ethics (Vol. 2, No. 11, p. 10).

Geurs, K. T., Patuelli, R., & Dentinho, T. P. (Eds.). (2016). Accessibility, Equity and Efficiency: Challenges for transport and public services. Edward Elgar Publishing.

Gilio-Whitaker, D. (2019). *As long as grass grows: The Indigenous fight for environmental justice, from colonization to Standing Rock*. Beacon Press.

Gooch, D., Kelly, R. M., Stiver, A., van der Linden, J., Petre, M., Richards, M., ... & Walton, C. (2020). The benefits and challenges of using crowdfunding to facilitate community-led projects in the context of digital civics. *International Journal of Human-Computer Studies*, 134, 33-43.

Gray, K. Build the Crowd: The Changing World of Public Infrastructure, (2013). http://www.wired.co.uk/magazine/archive/2013/11/features/built-by-the-crowd

Hickey, P. J., Erfani, A., & Cui, Q. (2022). Use of LinkedIn Data and Machine Learning to Analyze Gender Differences in Construction Career Paths. *Journal of Management in Engineering*, 38(6), 04022060.

Hopkins, E., Potoglou, D., Orford, S., & Cipcigan, L. (2023). Can the equitable roll out of electric vehicle charging infrastructure be achieved? *Renewable and Sustainable Energy Reviews*, 182, 113398.

Hsu, C. W., & Fingerman, K. (2021). Public electric vehicle charger access disparities across race and income in California. *Transport Policy*, 100, 59-67.

Kchaou-Boujelben, M. (2021). Charging station location problem: A comprehensive review on models and solution approaches. *Transportation Research Part C: Emerging Technologies*, 132, 103376.

Khan, H. A. U., Price, S., Avraam, C., & Dvorkin, Y. (2022). Inequitable access to EV charging infrastructure. The Electricity Journal, 35(3), 107096.

Kirby, E., & Worner, S. (2014). Crowd-funding: An infant industry growing fast. IOSCO Research Department, 2014, 1-63.

Kojola, E., & Pellow, D. N. (2021). New directions in environmental justice studies: examining the state and violence. *Environmental Politics*, *30*(1-2), 100-118.

Lam, P. T., & Law, A. O. (2016). Crowdfunding for renewable and sustainable energy projects: An exploratory case study approach. *Renewable and sustainable energy reviews*, 60, 11-20.

Lane, H., Sarkies, M., Martin, J., & Haines, T. (2017). Equity in healthcare resource allocation decision making: a systematic review. *Social science & medicine*, 175, 11-27.

Law, M., & Roy, A. (2021). A geospatial data fusion framework to quantify variations in electric vehicle charging demand. In Proceedings of the 4th ACM SIGSPATIAL international workshop on advances in resilient and intelligent cities (pp. 23-26).

Lebrouhi, B. E., Khattari, Y., Lamrani, B., Maaroufi, M., Zeraouli, Y., & Kousksou, T. (2021). Key challenges for a large-scale development of battery electric vehicles: A comprehensive review. *Journal of Energy Storage*, 44, 103273.

Ledna, C., Muratori, M., Brooker, A., Wood, E., & Greene, D. (2022). How to support EV adoption: Tradeoffs between charging infrastructure investments and vehicle subsidies in California. Energy Policy, 165, 112931.

Leuenberger, D. Z., & Lutte, R. (2022). Sustainability, gender equity, and air transport: Planning a stronger future. *Public Works Management & Policy*, 27(3), 238-251.

Li, G., Luo, T., & Song, Y. (2022). Spatial equity analysis of urban public services for electric vehicle charging—Implications of Chinese cities. *Sustainable Cities and Society*, *76*, 103519.

Li, S., Huang, Y., & Mason, S. J. (2016). A multi-period optimization model for the deployment of public electric vehicle charging stations on network. *Transportation Research Part C: Emerging Technologies*, 65, 128-143.

Li, Y., Zhang, F., & Ji, W. (2023). Integrated Data-Driven and Equity-Centered Framework for Highway Restoration Following Flood Inundation. *Journal of Management in Engineering*, *39*(3), 04023012.

Logue, D., & Grimes, M. (2022). Platforms for the people: Enabling civic crowdfunding through the cultivation of institutional infrastructure. Strategic Management Journal, 43(3), 663-693.

Lu, Y., Chang, R., & Lim, S. (2018). Crowdfunding for solar photovoltaics development: A review and forecast. *Renewable and sustainable energy reviews*, 93, 439-450.

Mandolakani, F. S., & Singleton, P. A. (2024). Electric vehicle charging infrastructure deployment: A discussion of equity and justice theories and accessibility measurement. *Transportation Research Interdisciplinary Perspectives*, 24, 101072.

Mastrangelo, L., Cruz-Ros, S., & Miquel-Romero, M. J. (2020). Crowdfunding success: The role of co-creation, feedback, and corporate social responsibility. *International Journal of Entrepreneurial Behavior & Research*, 26(3), 449-466.
28

Metais, M. O., Jouini, O., Perez, Y., Berrada, J., & Suomalainen, E. (2022). Too much or not enough? Planning electric vehicle charging infrastructure: A review of modeling options. *Renewable and Sustainable Energy Reviews*, 153, 111719.

Min, Y., & Lee, H. W. (2020). Social equity of clean energy policies in electric-vehicle charging infrastructure systems. In Construction Research Congress 2020 (pp. 221-229). Reston, VA: American Society of Civil Engineers.

Montgomery County. (2021). Montgomery County Climate Action Plan: Building a Healthy, Equitable, Resilient Community. < https://www.montgomerycountymd.gov/climate/index.html.>

Moriarty, P. (2022). Electric vehicles can have only a minor role in reducing transport's energy and environmental challenges. *AIMS Energy*, *10*(1), 131-148.

Nahmias-Biran, B. H., Sharaby, N., & Shiftan, Y. (2014). Equity aspects in transportation projects: Case study of transit fare change in Haifa. International Journal of Sustainable Transportation, 8(1), 69-83.

Nan, L., Tang, C., Wang, X., & Zhang, G. (2019). The real effects of transparency in crowdfunding. *Contemporary Accounting Research*.

Nazari-Heris, M., Loni, A., Asadi, S., & Mohammadi-ivatloo, B. (2022). Toward social equity access and mobile charging stations for electric vehicles: A case study in Los Angeles. *Applied Energy*, 311, 118704.

Nicholas, M. (2019). Estimating electric vehicle charging infrastructure costs across major US metropolitan areas. *International Council on Clean Transportation*, 14(11).

Nicholas, M., Hall, D., & Lutsey, N. (2019). Quantifying the electric vehicle charging infrastructure gap across US markets. Int. Counc. Clean Transp, 20, 1-39.

Nigro, N. (2019). EV Charging Financial Analysis Tool. *Atlas Public Policy*. <https://atlaspolicy.com/ev-charging-financial-analysis-tool/>.

Panah, P. G., Bornapour, S. M., Nosratabadi, S. M., & Guerrero, J. M. (2022). Hesitant fuzzy for conflicting criteria in multi-objective deployment of electric vehicle charging stations. *Sustainable Cities and Society*, *85*, 104054.

Pan, M. M., Uddin, M., & Lim, H. (2024). Understanding electric vehicle ownership using data fusion and spatial modeling. *Transportation Research Part D: Transport and Environment*, *127*, 104075.

Pan, S., Fulton, L. M., Roy, A., Jung, J., Choi, Y., & Gao, H. O. (2021). Shared use of electric autonomous vehicles: Air quality and health impacts of future mobility in the United States. *Renewable and Sustainable Energy Reviews*, *149*, 111380.

Papajohn, D., Cui, Q., & Bayraktar, M. E. (2011). Public-private partnerships in US transportation: Research overview and a path forward. *Journal of Management in Engineering*, *27*(3), 126-135.

Pranata, N., Firdaus, N., Mychelisda, E., & Hidayatina, A. (2022). Crowdfunding for infrastructure project financing: lessons learned for asian countries. Handbook of Research on Big Data, Green Growth, and Technology Disruption in Asian Companies and Societies, 276-300.

Roy, A., & Law, M. (2022). Examining spatial disparities in electric vehicle charging station placements using machine learning. *Sustainable cities and society*, 83, 103978.

Samarah, W. E. A. R., & Alkhatib, S. F. S. (2020). Crowdfunding operations: Outreach factors in developing economies. *Journal of Public Affairs*, 20(1), e1988.

Satterfield, C., & Nigro, N. (2020). A Financial Analysis of Common EV Charging Business Models for Retail Site Hosts. *Technical Report, Atlas Public Policy*.

Solaymani, S. (2019). CO2 emissions patterns in 7 top carbon emitter economies: The case of transport sector. *Energy*, 168, 989-1001.

Soliz, A., Carvalho, T., Sarmiento-Casas, C., Sánchez-Rodríguez, J., & El-Geneidy, A. (2023). Scaling up active transportation across North America: A comparative content analysis of policies through a social equity framework. *Transportation Research Part A: Policy and Practice*, 176, 103788.

Sotolongo, M. (2023). Defining environmental justice communities: Evaluating digital infrastructure in Southeastern states for Justice40 benefits allocation. *Applied Geography*, *158*, 103057.

Sperling, D. (2018). Three revolutions: Steering automated, shared, and electric vehicles to a better future. *Island Press*.

Sun, D., Kyere, F., Sampene, A. K., Asante, D., & Kumah, N. Y. G. (2023). An investigation on the role of electric vehicles in alleviating environmental pollution: evidence from five leading economies. *Environmental Science and Pollution Research*, 30(7), 18244-18259.

United States Department of Transportation. (2022). Equity action plan. https://www.transportation.gov/priorities/equity/actionplan

United States Department of Transportation. (2023). Background and Context for Urban Electric Mobility.

United States Environmental Protection Agency (EPA). (2023). Carbon Pollution from Transportation.

US Department of Transportation. (2023) Equitable Transportation Community Explorer, < https://experience.arcgis.com/experience/0920984aa80a4362b8778d779b090723/page/ETC-Explorer---National-Results>

Vulkan, N., Åstebro, T., & Sierra, M. F. (2016). Equity crowdfunding: A new phenomena. Journal of Business Venturing Insights, 5, 37-49.

Wang, Y., Apte, J. S., Hill, J. D., Ivey, C. E., Johnson, D., Min, E., ... & Marshall, J. D. (2023). Air quality policy should quantify effects on disparities. *Science*, *381*(6655), 272-274.

Wang, Y., Chen, J., & Shen, C. (2024). Exploring Online Perceptions of Justice in Large-Scale Infrastructure Projects: Temporal Patterns, Sentiment Characteristics, and Topic Changes. *Journal of Management in Engineering*, *40*(1), 05023008.

White House, T. (2022). Justice40: A whole-of-government initiative. Environmental justice. https://www.whitehouse.gov/environmentaljustice/justice40.

Yi, T., Cheng, X. B., Zheng, H., & Liu, J. P. (2019). Research on location and capacity optimization method for electric vehicle charging stations considering user's comprehensive satisfaction. *Energies*, 12(10), 1915.

Young, S. D., B. Mallory, and G. McCarthy. (2021). Interim implementation guidance for the Justice40 initiative. https://www.whitehouse.gov/wp-content/uploads/2021/07/M-21-28.pdf.

Zhang, R., & Fujimori, S. (2020). The role of transport electrification in global climate change mitigation scenarios. *Environmental Research Letters*, 15(3), 034019.

Zhao, D., Thakur, N., & Chen, J. (2020). Optimal design of energy storage system to buffer charging infrastructure in smart cities. *Journal of Management in Engineering*, 36(2), 04019048.

Zhu, L., Zhang, Q., Lu, H., Li, H., Li, Y., McLellan, B., & Pan, X. (2017). Study on crowdfunding's promoting effect on the expansion of electric vehicle charging piles based on game theory analysis. *Applied energy*, 196, 238-248.

Table 1- EV charging network cost items.

| Item | Cost |
|---|---|
| Level 3 Supercharges -350 kw | 40 units |
| Labor | $1,000,000 |
| Material | $1,000,000 |
| Permit & tax | $20,000 |
| Land | $1,000,000 |
| Level 2 Chargers | 60 units |
| Cost per charger- $5,000 | $300,000 |
| Traffic Management | $2,000,000 |
| Design and Construction | $800,000 |
| Public Engagement | $1,000,000 |
| Operation for 5-yr | $2,500,000 |
| Contingency | $380,000 |
| **Total** | **$10,000,000** |

Table 2- Return on investment.

|  | Year 1 to 5 | Year 1 to 5 | Year 6 to 10 | Year 6 to 10 |
|---|---|---|---|---|
| Revenue of level 3 charges | $360,000 | $360,000 | $360,000 | $360,000 |
| Revenue of level 2 charges | $360,000 | $360,000 | $360,000 | $360,000 |
| Total Revenue | $720,000 | $720,000 | $720,000 | $720,000 |
| Operation Cost | - | - | $500,000 | $500,000 |
| Operating Profit | $720,000 | $720,000 | $220,000 | $220,000 |
| Dividend to PPP Operator | $720,000 | - | $220,000 | - |
| Dividend to CPPP Operator | - | $360,000 | - | $110,000 |
| Dividend to Crowdfunding | - | $360,000 | - | $110,000 |
| **Return per share (crowdfunding)** | **-** | **$72** | **-** | **$22** |

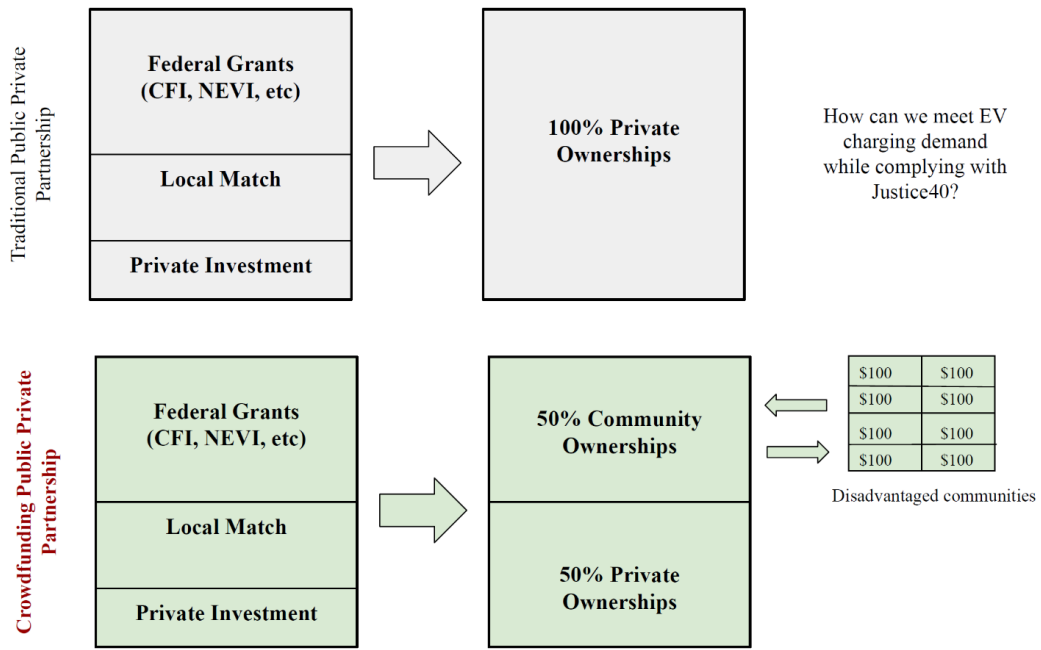

Figure 1. A Framework of the Crowdfunding for EV Infrastructure Pilot Program

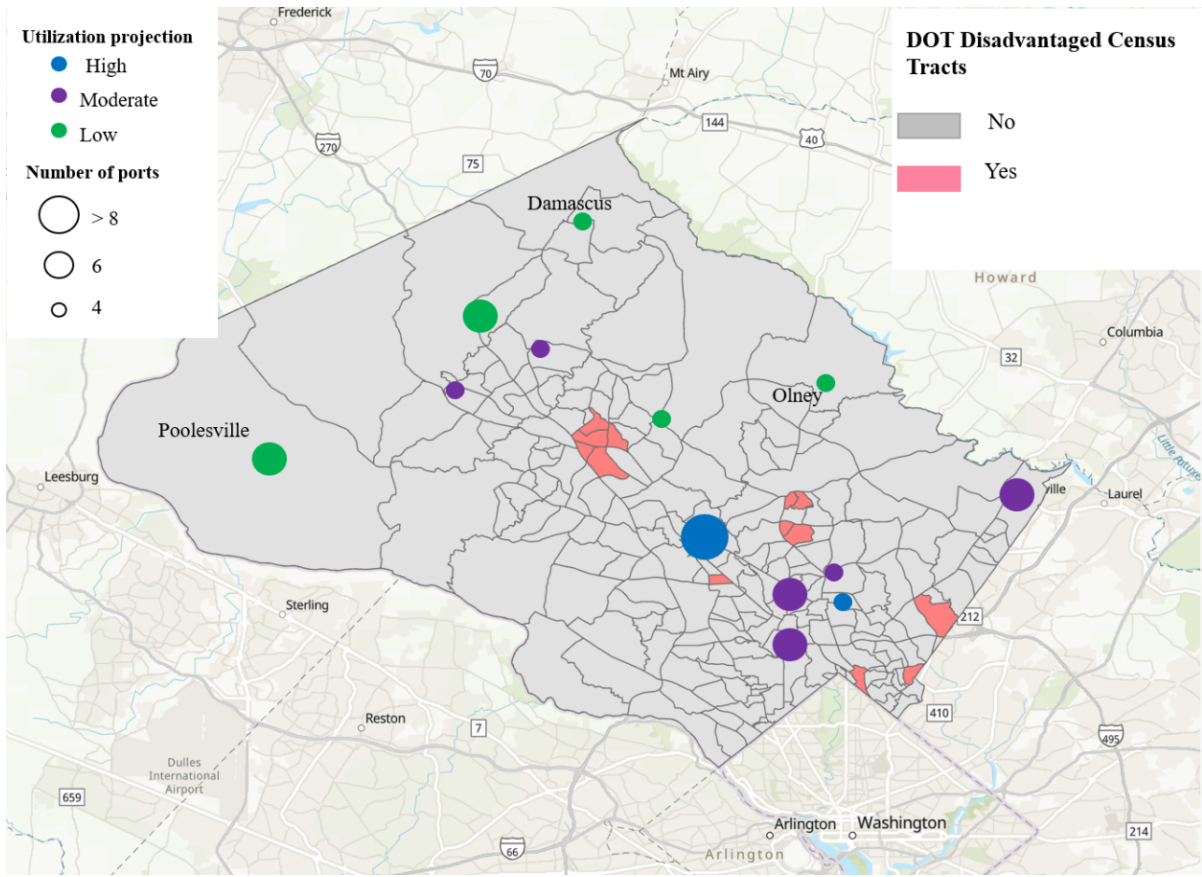

Figure 2. EV charging network locations and disadvantaged communities using ETC tool.